%% file: ak_xdmod_on_meltdown_and_spectre.tex
\documentclass[journal]{IEEEtran}

\usepackage{booktabs}

\usepackage{amsmath}
\usepackage{lscape}
\usepackage{authblk}
\usepackage{siunitx}

\usepackage{graphicx}
\usepackage{array}
\newcolumntype{H}{>{\setbox0=\hbox\bgroup}c<{\egroup}@{}} 
\usepackage{ragged2e}
\newcolumntype{R}[1]{>{\RaggedLeft\arraybackslash}p{#1}} 
\newcolumntype{L}[1]{>{\RaggedRight\arraybackslash}p{#1}}
\usepackage[flushleft]{threeparttable} 
\usepackage{multirow}

\usepackage{longtable}

\usepackage{arydshln}
\usepackage[letterpaper,margin=1in,marginparwidth=0.75in]{geometry}
\usepackage{float} 

\usepackage[backend=biber,style=nature,sorting=none,maxbibnames=99]{biblatex}
\begin{document}

\title{Effect of Meltdown and Spectre Patches on the Performance of HPC Applications}

\author{Nikolay~A.~Simakov,
        Martins~D.~Innus,
        Matthew~D.~Jones,
        Joseph~P.~White, Steven~M.~Gallo, 
        Robert~L.~DeLeon~and~Thomas~R.~Furlani
\thanks{Center for Computational Research, State University of New York, University at Buffalo, Buffalo, NY}%
\thanks{nikolays@buffalo.edu}}

\markboth{Preprint, January~2018}%
{AK XDMoD: Performance Effects of Meltdown and Spectre}

\maketitle

\begin{abstract}
In this work we examine how the updates addressing Meltdown and Spectre vulnerabilities impact the performance of HPC applications. To study this we use the  application kernel module of XDMoD to test the performance before and after the application of the vulnerability patches. We tested the performance difference for multiple application and benchmarks including: NWChem, NAMD, HPCC, IOR, MDTest and IMB. The results show that although some specific functions can have perfomance decreased by as much as 74\%, the majority of individual metrics indicates little to no decrease in performance. The real-world applications show a 2-3\% decrease in performance for single node jobs and a 5-11\% decrease for parallel multi node jobs.
\end{abstract}

\begin{IEEEkeywords}
HPC, Security, Performance
\end{IEEEkeywords}


\section{Introduction}
The recently discovered Meltdown~\cite{Lipp2018meltdown} and Spectre~\cite{Kocher2018spectre} vulnerabilities allow reading of process memory by other unauthorized processes. This poses a significant security risk on multi-user platforms including HPC resources that can result in the compromise of proprietary or sensitive information ~\cite{Lipp2018meltdown,Kocher2018spectre}.  Software patches released to mitigate the security vulnerabilities have the potential to significantly impact performance.  According to Redhat ~\cite{RedhatPerfImpacts2018} Linux OS remedies can degrade performance overall by 1-20\%. In order to quantify the impact, particularly on HPC applications, we performed independent tests utilizing XDMoD's application kernel capability ~\cite{Simakov:2015:AppKernels}.

\begin{figure}[h]
\centering
\includegraphics[width=3.25in]{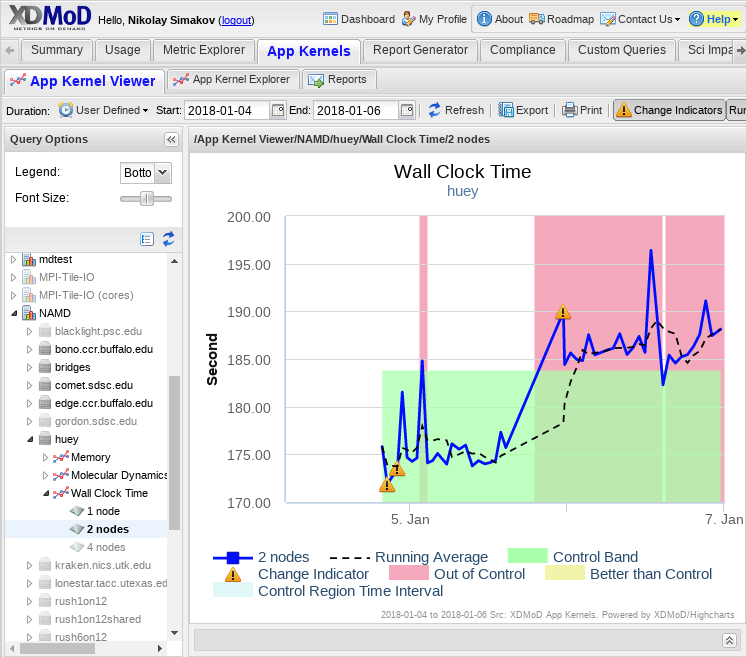}
\caption{Screen-shot of the application kernel performance monitoring module of XDMod showing the change in performance of NAMD executed on 2-nodes.   The module automatically calculates control regions and performs an automatic detection of performance degradation and application environment changes.}
\label{fig:xdmod_ak_screenshot}
\end{figure}

The XD Metrics on Demand (XDMoD) tool, which is designed for the comprehensive management of HPC systems, provides users, managers, and operations staff with access to   utilization data, job and system level performance data, and quality of service data for HPC resources~\cite{furlani2013using}. Originally developed to provide independent audit capability for the XSEDE program, XDMoD was later open-sourced and is widely used by university, government, and industry HPC centers~\cite{Palmer:2015}. The application kernel performance monitoring module of XDMoD ~\cite{Simakov:2015:AppKernels} allows automatic performance monitoring of HPC resources through the periodic execution of application kernels, which are based on benchmarks or real-world applications implemented with sensible input parameters (see Figure \ref{fig:xdmod_ak_screenshot} for web interface screen-shot).

Since the application kernels, which are computationally lightweight, are designed to run continuously on a given HPC system, they are ideal for detecting differences in application performance when system wide changes (hardware or software) are made. Accordingly, XDMoD's application kernels were employed here to determine if the software patches that mitigate the Meltdown and Spectre vulnerabilities significantly impact performance.

\begin{table*}[t]
\begin{threeparttable}
\centering
\caption{Change in walltime upon patch application.}
\label{table:ak-walltime-difference}
\begin{tabular}{|l|R{0.4in}|R{0.4in}|R{0.6in}|R{0.4in}|R{0.4in}|R{0.4in}|R{0.4in}|R{0.4in}|R{0.4in}|}
\hline
\multirow{2}{*}{Application} & 
\multirow{2}{0.4in}{\centering Number of Nodes} & 
\multirow{2}{0.4in}{\centering Difference, \%\tnote{1}} & 
\multirow{2}{0.6in}{\centering Are the means different?\tnote{2}} & 
\multicolumn{3}{l|}{Before Patch Application} & 
\multicolumn{3}{l|}{After Patch Application} \\[0.1in] \cline{5-10} 
 &  &  &  & 
{\centering Mean, Seconds} & {\centering Standard Deviation, Seconds} & {\centering Number of Runs} & 
{\centering Mean, Seconds} & {\centering Standard Deviation, Seconds} & {\centering Number of Runs} \\[0.1in] \hline

NAMD & 1 & 3.3 & Y & 306.6 & 1.44 & 24 & 316.9 & 3.05 & 56 \\ \hline
NAMD & 2 & 6.9 & Y & 175.4 & 2.78 & 22 & 188.1 & 3.49 & 56 \\ \hline
NWChem & 1 & 2.6 & Y & 77.8 & 1.91 & 23 & 79.9 & 1.11 & 59 \\ \hline
NWChem & 2 & 10.7 & Y & 58.4 & 1.05 & 21 & 65.0 & 4.16 & 56 \\ \hline
HPCC & 1 & 2.2 & Y & 304.1 & 6.39 & 23 & 310.9 & 4.88 & 56 \\ \hline
HPCC & 2 & 5.3 & Y & 345.1 & 5.41 & 22 & 364.0 & 8.44 & 56 \\ \hline
IMB & 2 & 4 & Y & 14.8 & 0.54 & 21 & 15.4 & 1.39 & 56 \\ \hline
IOR & 1 & 3.9 & Y & 188.5 & 9.41 & 21 & 195.9 & 11.69 & 55 \\ \hline
IOR & 2 & 1.5 & N & 371.1 & 12.23 & 22 & 376.7 & 19.50 & 56 \\ \hline
IOR.local & 1 & 2.1 & N & 462.8 & 16.37 & 12 & 472.8 & 19.03 & 56 \\ \hline
MDTest & 1 & 21.5 & Y & 30.5 & 3.17 & 21 & 37.8 & 4.10 & 56 \\ \hline
MDTest & 2 & 9.3 & Y & 166.7 & 3.60 & 23 & 182.8 & 5.30 & 55 \\ \hline
MDTest.local & 1 & 56.4 & Y & 3.8 & 0.62 & 12 & 6.7 & 2.61 & 56 \\ \hline
\end{tabular}
\begin{tablenotes}
\item[1] Differences are calculated as the new mean value minus the old mean value divided by the average of the two means. A larger difference indicates poorer performance after the patch.
\item[2] The Welch two sample, two sided, t-test with $\alpha=0.5$ was used to determine if the before and after test results were drawn from distributions with statistically significantly different means.
\end{tablenotes}
\end{threeparttable}
\end{table*}

\section{Methods}
\subsection{Selected Application Kernels}

The following XDMoD application kernels were chosen for this test: NAMD~\cite{namd}, NWChem~\cite{nwchem}, HPC Challenge Benchmark suite (HPCC)~\cite{hpcc} (which includes
memory bandwidth micro-benchmark STREAM~\cite{stream} and the NASA parallel benchmarks (NPB)\cite{npb-web}), interconnect/MPI benchmarks (IMB)~\cite{osumpi,imb}, IOR~\cite{ior} and MDTest~\cite{mdtest}. The first two are based on widely used scientific applications and the others are based on commonly deployed benchmarks. Most of the application kernels were executed on one or two nodes, 8 and 16 cores respectively. For more details on application kernels refer to ~\cite{Simakov:2015:AppKernels}.

IOR and MDTest were executed on the parallel file system (GPFS) as well as the local file system.   In order to differentiate between the two file systems, we use a ".local" suffix in the reported results when the local file system is used (e.g. IOR.local).

\subsection{System}

The tests were performed on a development cluster at the Center for Computational Research (CCR), SUNY, University at Buffalo. The cluster consists of eight nodes (8-cores, 24GiB RAM) with two Intel L5520 CPUs connected by QDR Mellanox Infiniband. The nodes have access to a 3 PB IBM GPFS storage system shared with other HPC resources in CCR. The operating system is CentOS Linux release 7.4.1708.

\subsection{Patches}

To fix the Meltdown and Spectre vulnerabilities a new kernel was installed. Specifically,  kernel-3.10.0-693.5.2.el7.x86\_64 was updated with kernel-3.10.0-693.11.6.el7.x86\_64 which fixes CVE-2017-5753, CVE-2017-5715 and CVE-2017-5754 vulnerabilities.

\subsection{Comparison of the Results}

The tests were run prior to and after application of the vulnerability updates. The "before" tests include approximately 20 runs for most of the application kernels. The "after" tests include approximately 50 runs for all application kernels. The comparison of before and after distributions were determine using the Welch two sample, two sided, t-test with $\alpha$ parameter equal to 0.05. That is, we consider the means of two distributions to be different if the probability that such test results could be obtained from equal distributions is less than or equal to 0.05.

\begin{figure*}[h]
\centering
\includegraphics[width=5in]{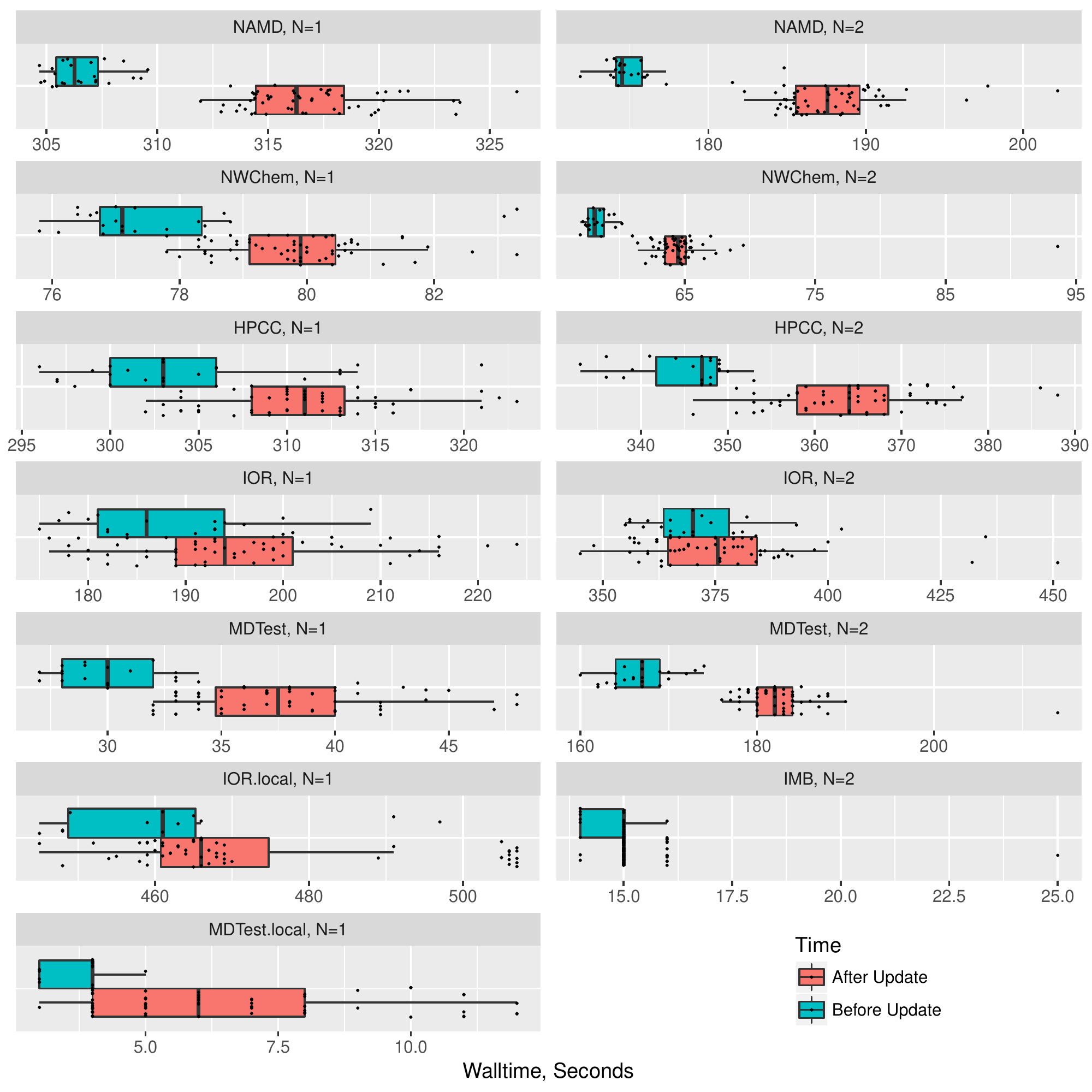}
\caption{\label{fig:walltime_change} Application kernel walltime comparisons before and after the updates. Box plot diagram is used to show sample statistics. Left side of the box, vertical line within the box and right side of the box show first quartile, medium and third quartile. In addition all measurements are plotted using round points.}
\end{figure*}

\section{Results and Discussion}

Table \ref{table:ak-walltime-difference} and Figure \ref{fig:walltime_change} show the change in walltime before and after the patches for the  suite of application kernels employed in this study. For the compute intensive applications (NAMD, NWChem and HPCC), the performance degradation is around 2-3\% for parallel single node jobs. However it increases to 5-11\%  for the case of two nodes.

IOR and MDTest benchmarks measure the performance of the file system.  As discussed in the introduction, we tested both the parallel and local file systems. Tables \ref{table:ior-metrics} and \ref{table:mdtest-metrics} show selected results for these tests. In both cases there is a significant decrease in performance for file meta-data operations (10-20\%). However, the performance degradation for read and write operations is only in the range of 0-3\%. Based on these findings, the performance degradation should be smaller for applications that use a small number of large files versus those that use a large number of small files. Data processing applications may therefore be particularly sensitive to the patches employed to mitigate the vulnerabilities. 

The IMB test shows that most reported metrics are degraded by more than by 2\% (Table \ref{table:imb-metrics}).

The HPCC benchmark performs various tests from linear algebra, fast Fourier transformation (FFT) and memory manipulation. Interestingly the simple arrays manipulations (STREAM tests: arrays addition, copying and scaling) are actually faster in case of two nodes (Table \ref{table:namd-nwchem-hpcc-metrics}). However FFT, matrix manipulation and matrix transposition get slower. The surprising performance improvement in STREAM tests might be due to other changes in kernel. Anyway this improvement does not transfer to matrix manipulation and matrix transposition, which are 2\% and 10\% slower (two nodes).



\section{Conclusions and Future Plans}

Some of the individually measured simple metrics show a significant decrease in performance, notably MPI random access, memory copying and file metadata operations. Many other metrics show little to no change.

Overall the compute intensive single node applications have a moderate decrease in the performance around 2-3\%. However, multi-node parallel jobs suffer a  5-11\% decrease in performance. This can Probably be addressed in the compiler and MPI libraries.

These tests were executed in a relatively isolated environment.  After the updates are applied on our production system we will perform additional tests with a larger number of nodes and for more application kernels.

\section{Acknowledgements}
We gratefully acknowledge the support of NSF awards OCI 1025159, ACI 1445806, and OCI 1203560. 

\printbibliography

\input{selected_metrics.tex}

\end{document}

%% file: selected_metrics.tex
\begin{table*}[!t]
\centering
\footnotesize
\caption{Changes in selected measured metrics from NAMD, NWChem and HPCC.}
\label{table:namd-nwchem-hpcc-metrics}
\begin{tabular}
{|L{0.1in}|L{0.4in}|R{0.2in}|L{1.2in}|R{0.3in}|R{0.3in}|R{0.4in}|R{0.4in}|R{0.2in}|R{0.4in}|R{0.4in}|R{0.2in}|L{0.6in}|}
\hline
\multirow{2}{*}{\#} & 
\multirow{2}{*}{Application} & 
\multirow{2}{0.2in}{Nodes} & 
\multirow{2}{1.0in}{Metric} & 
\multirow{2}{0.3in}{Diff., \%} & 
\multirow{2}{0.3in}{Distr. are Different} & 
\multicolumn{3}{l|}{Before Patch Application} & 
\multicolumn{3}{l|}{After Patch Application} &
\multirow{2}{*}{Units} \\[0.12in]  \cline{7-12} 
 &  &  &  &  &  & 
{\centering Mean, Seconds} & {\centering St.Dev., Seconds} & {Nruns} & 
{\centering Mean, Seconds} & {\centering St.Dev., Seconds} & {Nruns} & \\[0.12in] \hline

2 & NAMD & 1 & Molecular Dynamics Simulation Performance & -3.5 & TRUE & 7.67E-10 & 7.11E-12 & 24 & 7.40E-10 & 1.43E-11 & 56 & Second per Day \\ \hline
3 & NAMD & 1 & Wall Clock Time & 3.4 & TRUE & 3.07E+02 & 1.44E+00 & 24 & 3.17E+02 & 3.05E+00 & 56 & Second \\ \hline
5 & NAMD & 2 & Molecular Dynamics Simulation Performance & -6.6 & TRUE & 1.48E-09 & 5.60E-11 & 22 & 1.38E-09 & 8.44E-11 & 56 & Second per Day \\ \hline
6 & NAMD & 2 & Wall Clock Time & 7.2 & TRUE & 1.75E+02 & 2.78E+00 & 22 & 1.88E+02 & 3.49E+00 & 56 & Second \\ \hline
15 & NWChem & 1 & User Time & 1.1 & TRUE & 7.35E+01 & 7.36E-01 & 23 & 7.43E+01 & 1.08E+00 & 59 & Second \\ \hline
16 & NWChem & 1 & Wall Clock Time & 2.7 & TRUE & 7.78E+01 & 1.91E+00 & 23 & 7.99E+01 & 1.11E+00 & 59 & Second \\ \hline
25 & NWChem & 2 & User Time & 10.2 & TRUE & 4.75E+01 & 8.37E-01 & 21 & 5.23E+01 & 3.84E+00 & 56 & Second \\ \hline
26 & NWChem & 2 & Wall Clock Time & 11.3 & TRUE & 5.84E+01 & 1.05E+00 & 21 & 6.50E+01 & 4.16E+00 & 56 & Second \\ \hline
27 & HPCC & 1 & Matrix Multiplication (DGEMM) Floating-Point Performance & -1.2 & TRUE & 8.50E+03 & 5.66E+01 & 23 & 8.40E+03 & 1.11E+02 & 56 & MFLOP per Second \\ \hline
28 & HPCC & 1 & Average STREAM 'Add' Memory Bandwidth & 5 & FALSE & 3.17E+03 & 4.51E+02 & 23 & 3.33E+03 & 5.38E+02 & 56 & MByte per Second \\ \hline
29 & HPCC & 1 & Average STREAM 'Copy' Memory Bandwidth & 3.4 & FALSE & 4.36E+03 & 4.96E+02 & 23 & 4.51E+03 & 5.61E+02 & 56 & MByte per Second \\ \hline
30 & HPCC & 1 & Average STREAM 'Scale' Memory Bandwidth & 3.5 & FALSE & 2.95E+03 & 4.05E+02 & 23 & 3.05E+03 & 4.54E+02 & 56 & MByte per Second \\ \hline
31 & HPCC & 1 & Average STREAM 'Triad' Memory Bandwidth & 8.8 & TRUE & 3.29E+03 & 5.01E+02 & 23 & 3.58E+03 & 6.68E+02 & 56 & MByte per Second \\ \hline
32 & HPCC & 1 & Fast Fourier Transform (FFTW) Floating-Point Performance & -5.9 & TRUE & 7.93E+03 & 5.55E+02 & 23 & 7.46E+03 & 1.50E+03 & 56 & MFLOP per Second \\ \hline
33 & HPCC & 1 & High Performance LINPACK Efficiency & -5.4 & TRUE & 9.26E+01 & 7.86E+00 & 23 & 8.76E+01 & 6.19E+00 & 56 & Percent \\ \hline
34 & HPCC & 1 & High Performance LINPACK Floating-Point Performance & -4.1 & TRUE & 6.18E+04 & 1.24E+03 & 23 & 5.93E+04 & 1.24E+03 & 56 & MFLOP per Second \\ \hline
35 & HPCC & 1 & High Performance LINPACK Run Time & 4.3 & TRUE & 8.63E+01 & 1.88E+00 & 23 & 9.00E+01 & 1.97E+00 & 56 & Second \\ \hline
36 & HPCC & 1 & MPI Random Access & -23.3 & TRUE & 2.09E+00 & 8.59E-02 & 23 & 1.61E+00 & 3.78E-02 & 56 & MUpdate per Second \\ \hline
37 & HPCC & 1 & Parallel Matrix Transpose (PTRANS) & -12.3 & TRUE & 3.03E+03 & 4.37E+02 & 23 & 2.66E+03 & 8.46E+02 & 56 & MByte per Second \\ \hline
38 & HPCC & 1 & Wall Clock Time & 2.2 & TRUE & 3.04E+02 & 6.39E+00 & 23 & 3.11E+02 & 4.88E+00 & 56 & Second \\ \hline
39 & HPCC & 2 & Matrix Multiplication (DGEMM) Floating-Point Performance & -2 & TRUE & 8.53E+03 & 4.22E+01 & 22 & 8.36E+03 & 9.18E+01 & 56 & MFLOP per Second \\ \hline
40 & HPCC & 2 & Average STREAM 'Add' Memory Bandwidth & 14.1 & TRUE & 3.12E+03 & 2.91E+02 & 22 & 3.56E+03 & 6.40E+02 & 56 & MByte per Second \\ \hline
41 & HPCC & 2 & Average STREAM 'Copy' Memory Bandwidth & 10.9 & TRUE & 4.35E+03 & 3.16E+02 & 22 & 4.82E+03 & 6.27E+02 & 56 & MByte per Second \\ \hline
42 & HPCC & 2 & Average STREAM 'Scale' Memory Bandwidth & 13.8 & TRUE & 2.92E+03 & 3.09E+02 & 22 & 3.32E+03 & 5.95E+02 & 56 & MByte per Second \\ \hline
43 & HPCC & 2 & Average STREAM 'Triad' Memory Bandwidth & 16.5 & TRUE & 3.21E+03 & 3.28E+02 & 22 & 3.75E+03 & 6.94E+02 & 56 & MByte per Second \\ \hline
44 & HPCC & 2 & Fast Fourier Transform (FFTW) Floating-Point Performance & -6.4 & TRUE & 1.23E+04 & 6.35E+02 & 22 & 1.16E+04 & 1.65E+03 & 56 & MFLOP per Second \\ \hline
45 & HPCC & 2 & High Performance LINPACK Efficiency & -12.4 & TRUE & 9.62E+01 & 1.01E+01 & 22 & 8.42E+01 & 7.16E+00 & 56 & Percent \\ \hline
46 & HPCC & 2 & High Performance LINPACK Floating-Point Performance & -8.6 & TRUE & 1.22E+05 & 4.85E+02 & 22 & 1.11E+05 & 2.75E+03 & 56 & MFLOP per Second \\ \hline
47 & HPCC & 2 & High Performance LINPACK Run Time & 9.5 & TRUE & 1.24E+02 & 4.94E-01 & 22 & 1.35E+02 & 3.42E+00 & 56 & Second \\ \hline
48 & HPCC & 2 & MPI Random Access & -54 & TRUE & 9.59E+00 & 3.29E-01 & 22 & 4.41E+00 & 2.69E-01 & 56 & MUpdate per Second \\ \hline
49 & HPCC & 2 & Parallel Matrix Transpose (PTRANS) & -9.6 & TRUE & 2.62E+03 & 1.87E+02 & 22 & 2.36E+03 & 3.44E+02 & 56 & MByte per Second \\ \hline
50 & HPCC & 2 & Wall Clock Time & 5.5 & TRUE & 3.45E+02 & 5.41E+00 & 22 & 3.64E+02 & 8.44E+00 & 56 & Second \\ \hline
\end{tabular}
\end{table*}

\begin{table*}[!t]
\centering
\footnotesize
\caption{Changes in selected measured metrics from IMB.}
\label{table:imb-metrics}
\begin{tabular}
{|L{0.1in}|L{0.4in}|R{0.2in}|L{1.2in}|R{0.3in}|R{0.3in}|R{0.4in}|R{0.4in}|R{0.2in}|R{0.4in}|R{0.4in}|R{0.2in}|L{0.6in}|}
\hline
\multirow{2}{*}{\#} & 
\multirow{2}{*}{Application} & 
\multirow{2}{0.2in}{Nodes} & 
\multirow{2}{1.0in}{Metric} & 
\multirow{2}{0.3in}{Diff., \%} & 
\multirow{2}{0.3in}{Distr. are Different} & 
\multicolumn{3}{l|}{Before Patch Application} & 
\multicolumn{3}{l|}{After Patch Application} &
\multirow{2}{*}{Units} \\[0.12in]  \cline{7-12} 
 &  &  &  &  &  & 
{\centering Mean, Seconds} & {\centering St.Dev., Seconds} & {Nruns} & 
{\centering Mean, Seconds} & {\centering St.Dev., Seconds} & {Nruns} & \\[0.12in] \hline

51 & IMB & 2 & Max Exchange Bandwidth & 0.2 & FALSE & 3.90E+03 & 3.60E+01 & 21 & 3.91E+03 & 6.30E+01 & 56 & MByte per Second \\ \hline
52 & IMB & 2 & Max MPI-2 Bidirectional 'Get' Bandwidth (aggregate) & -0.7 & FALSE & 1.99E+03 & 4.31E+01 & 21 & 1.98E+03 & 5.40E+01 & 56 & MByte per Second \\ \hline
53 & IMB & 2 & Max MPI-2 Bidirectional 'Get' Bandwidth (non-aggregate) & -2.8 & TRUE & 2.11E+03 & 5.71E+01 & 21 & 2.05E+03 & 4.68E+01 & 56 & MByte per Second \\ \hline
54 & IMB & 2 & Max MPI-2 Bidirectional 'Put' Bandwidth (aggregate) & 0.4 & FALSE & 2.05E+03 & 6.51E+01 & 21 & 2.06E+03 & 3.02E+01 & 56 & MByte per Second \\ \hline
55 & IMB & 2 & Max MPI-2 Bidirectional 'Put' Bandwidth (non-aggregate) & -1.7 & TRUE & 2.12E+03 & 3.61E+01 & 21 & 2.08E+03 & 4.37E+01 & 56 & MByte per Second \\ \hline
56 & IMB & 2 & Max MPI-2 Unidirectional 'Get' Bandwidth (aggregate) & -0.1 & FALSE & 3.10E+03 & 1.45E+01 & 21 & 3.10E+03 & 8.30E+00 & 56 & MByte per Second \\ \hline
57 & IMB & 2 & Max MPI-2 Unidirectional 'Get' Bandwidth (non-aggregate) & -1 & TRUE & 2.93E+03 & 5.45E+01 & 21 & 2.90E+03 & 5.59E+01 & 56 & MByte per Second \\ \hline
58 & IMB & 2 & Max MPI-2 Unidirectional 'Put' Bandwidth (aggregate) & 0 & FALSE & 3.12E+03 & 6.18E+00 & 21 & 3.12E+03 & 7.70E+00 & 56 & MByte per Second \\ \hline
59 & IMB & 2 & Max MPI-2 Unidirectional 'Put' Bandwidth (non-aggregate) & -1.1 & TRUE & 2.96E+03 & 4.11E+01 & 21 & 2.92E+03 & 5.11E+01 & 56 & MByte per Second \\ \hline
60 & IMB & 2 & Max PingPing Bandwidth & -0.1 & FALSE & 2.60E+03 & 3.43E+01 & 21 & 2.60E+03 & 3.00E+01 & 56 & MByte per Second \\ \hline
61 & IMB & 2 & Max PingPong Bandwidth & -0.1 & FALSE & 3.08E+03 & 9.99E+00 & 21 & 3.08E+03 & 7.15E+00 & 56 & MByte per Second \\ \hline
62 & IMB & 2 & Max SendRecv Bandwidth & -0.4 & FALSE & 5.22E+03 & 7.35E+01 & 21 & 5.20E+03 & 7.24E+01 & 56 & MByte per Second \\ \hline
63 & IMB & 2 & Min AllGather Latency & 1.5 & TRUE & 2.54E-06 & 5.12E-08 & 21 & 2.58E-06 & 6.88E-08 & 56 & Second \\ \hline
64 & IMB & 2 & Min AllGatherV Latency & -1 & FALSE & 2.99E-06 & 1.16E-07 & 21 & 2.96E-06 & 6.92E-08 & 56 & Second \\ \hline
65 & IMB & 2 & Min AllReduce Latency & -2.3 & TRUE & 3.02E-06 & 1.22E-07 & 21 & 2.95E-06 & 7.10E-08 & 56 & Second \\ \hline
66 & IMB & 2 & Min AllToAll Latency & 0.9 & TRUE & 2.53E-06 & 3.66E-08 & 21 & 2.55E-06 & 3.56E-08 & 56 & Second \\ \hline
67 & IMB & 2 & Min AllToAllV Latency & 3.8 & TRUE & 3.21E-06 & 3.01E-08 & 21 & 3.33E-06 & 1.14E-07 & 56 & Second \\ \hline
68 & IMB & 2 & Min Barrier Latency & 3.3 & TRUE & 2.41E-06 & 8.48E-08 & 21 & 2.48E-06 & 1.24E-07 & 56 & Second \\ \hline
69 & IMB & 2 & Min Broadcast Latency & -0.7 & FALSE & 2.40E-06 & 2.53E-08 & 21 & 2.39E-06 & 8.71E-08 & 56 & Second \\ \hline
70 & IMB & 2 & Min Gather Latency & -0.4 & FALSE & 2.60E-06 & 9.05E-08 & 21 & 2.59E-06 & 3.39E-08 & 56 & Second \\ \hline
71 & IMB & 2 & Min GatherV Latency & 0.9 & FALSE & 2.44E-06 & 2.10E-08 & 21 & 2.46E-06 & 8.03E-08 & 56 & Second \\ \hline
72 & IMB & 2 & Min MPI-2 'Accumulate' Latency (aggregate) & 0.5 & FALSE & 1.00E-06 & 7.08E-08 & 21 & 1.01E-06 & 6.05E-08 & 56 & Second \\ \hline
73 & IMB & 2 & Min MPI-2 'Accumulate' Latency (non-aggregate) & 1.8 & FALSE & 6.29E-06 & 2.50E-07 & 21 & 6.40E-06 & 3.00E-07 & 56 & Second \\ \hline
74 & IMB & 2 & Min MPI-2 Window Creation Latency & 0.3 & FALSE & 2.44E-05 & 1.36E-07 & 21 & 2.45E-05 & 2.06E-07 & 56 & Second \\ \hline
75 & IMB & 2 & Min Reduce Latency & 3.5 & TRUE & 2.74E-06 & 6.69E-08 & 21 & 2.84E-06 & 1.05E-07 & 56 & Second \\ \hline
76 & IMB & 2 & Min ReduceScatter Latency & 0.8 & FALSE & 1.65E-06 & 8.98E-08 & 21 & 1.67E-06 & 1.03E-07 & 56 & Second \\ \hline
77 & IMB & 2 & Min Scatter Latency & 1.6 & TRUE & 2.58E-06 & 1.60E-08 & 21 & 2.63E-06 & 5.97E-08 & 56 & Second \\ \hline
78 & IMB & 2 & Min ScatterV Latency & 0.1 & FALSE & 2.54E-06 & 7.85E-08 & 21 & 2.54E-06 & 4.09E-08 & 56 & Second \\ \hline
79 & IMB & 2 & Wall Clock Time & 4 & TRUE & 1.48E+01 & 5.39E-01 & 21 & 1.54E+01 & 1.39E+00 & 56 & Second \\ \hline
\end{tabular}
\end{table*}

\begin{table*}[!t]
\centering
\footnotesize
\caption{Changes in all measured metrics from IOR}
\label{table:ior-metrics}
\begin{tabular}
{|L{0.1in}|L{0.6in}|R{0.2in}|L{1.0in}|R{0.3in}|R{0.3in}|R{0.4in}|R{0.4in}|R{0.2in}|R{0.4in}|R{0.4in}|R{0.2in}|L{0.6in}|}
\hline
\multirow{2}{*}{\#} & 
\multirow{2}{*}{Application} & 
\multirow{2}{0.2in}{Nodes} & 
\multirow{2}{1.0in}{Metric} & 
\multirow{2}{0.3in}{Diff., \%} & 
\multirow{2}{0.3in}{Distr. Can be Distin- gui- shed?} & 
\multicolumn{3}{l|}{Before Patch Application} & 
\multicolumn{3}{l|}{After Patch Application} &
\multirow{2}{*}{Units} \\[0.1in]  \cline{7-12} 
 &  &  &  &  &  & 
{\centering Mean, Seconds} &l {\centering Standard Deviation, Seconds} & {Nruns} & 
{\centering Mean, Seconds} & {\centering Standard Deviation, Seconds} & {Nruns} & \\[0.1in] \hline
 
88 & IOR & 1 & HDF5 Independent N-to-1  File Open Time (Read) & 36.9 & FALSE & 2.13E-01 & 1.30E-01 & 21 & 2.92E-01 & 2.88E-01 & 55 & Second \\ \hline
89 & IOR & 1 & HDF5 Independent N-to-1  File Open Time (Write) & 0.5 & FALSE & 1.65E-01 & 2.06E-01 & 21 & 1.66E-01 & 2.12E-01 & 55 & Second \\ \hline
90 & IOR & 1 & HDF5 Independent N-to-1 Read Aggregate Throughput & -6.3 & FALSE & 1.64E+02 & 2.17E+01 & 21 & 1.53E+02 & 2.04E+01 & 55 & MByte per Second \\ \hline
91 & IOR & 1 & HDF5 Independent N-to-1 Write Aggregate Throughput & -3.7 & FALSE & 1.66E+02 & 2.67E+01 & 21 & 1.60E+02 & 2.26E+01 & 55 & MByte per Second \\ \hline
138 & IOR & 1 & POSIX N-to-N  File Open Time (Read) & 1.2 & FALSE & 1.44E+00 & 1.77E+00 & 21 & 1.45E+00 & 1.75E+00 & 55 & Second \\ \hline
139 & IOR & 1 & POSIX N-to-N  File Open Time (Write) & -28.9 & TRUE & 4.41E+00 & 2.25E+00 & 21 & 3.14E+00 & 1.99E+00 & 55 & Second \\ \hline
140 & IOR & 1 & POSIX N-to-N Read Aggregate Throughput & 0 & FALSE & 2.32E+02 & 5.47E+00 & 21 & 2.32E+02 & 6.43E+00 & 55 & MByte per Second \\ \hline
141 & IOR & 1 & POSIX N-to-N Write Aggregate Throughput & 14.6 & TRUE & 2.33E+02 & 4.37E+01 & 21 & 2.67E+02 & 3.47E+01 & 55 & MByte per Second \\ \hline
142 & IOR & 1 & Wall Clock Time & 3.9 & TRUE & 1.88E+02 & 9.41E+00 & 21 & 1.96E+02 & 1.17E+01 & 55 & Second \\ \hline
151 & IOR & 2 & HDF5 Independent N-to-1  File Open Time (Read) & -18.2 & FALSE & 7.01E-01 & 1.10E+00 & 22 & 5.74E-01 & 6.34E-01 & 56 & Second \\ \hline
152 & IOR & 2 & HDF5 Independent N-to-1  File Open Time (Write) & 4.1 & FALSE & 2.24E-01 & 1.35E-01 & 22 & 2.33E-01 & 1.86E-01 & 56 & Second \\ \hline
153 & IOR & 2 & HDF5 Independent N-to-1 Read Aggregate Throughput & -1.9 & FALSE & 1.95E+02 & 2.29E+01 & 22 & 1.92E+02 & 2.06E+01 & 56 & MByte per Second \\ \hline
154 & IOR & 2 & HDF5 Independent N-to-1 Write Aggregate Throughput & -3.6 & FALSE & 1.59E+02 & 2.48E+01 & 22 & 1.54E+02 & 2.28E+01 & 56 & MByte per Second \\ \hline
201 & IOR & 2 & POSIX N-to-N  File Open Time (Read) & 26.6 & FALSE & 3.06E+00 & 3.08E+00 & 22 & 3.88E+00 & 3.06E+00 & 56 & Second \\ \hline
202 & IOR & 2 & POSIX N-to-N  File Open Time (Write) & -20 & FALSE & 8.43E+00 & 3.29E+00 & 22 & 6.75E+00 & 3.92E+00 & 56 & Second \\ \hline
203 & IOR & 2 & POSIX N-to-N Read Aggregate Throughput & 0.2 & FALSE & 2.67E+02 & 3.25E+00 & 22 & 2.68E+02 & 2.43E+00 & 56 & MByte per Second \\ \hline
204 & IOR & 2 & POSIX N-to-N Write Aggregate Throughput & 3.2 & FALSE & 2.29E+02 & 3.33E+01 & 22 & 2.36E+02 & 3.17E+01 & 56 & MByte per Second \\ \hline
205 & IOR & 2 & Wall Clock Time & 1.5 & FALSE & 3.71E+02 & 1.22E+01 & 22 & 3.77E+02 & 1.95E+01 & 56 & Second \\ \hline
214 & IOR.local & 1 & HDF5 Independent N-to-1  File Open Time (Read) & 7.6 & TRUE & 5.57E-02 & 3.66E-03 & 12 & 5.99E-02 & 4.41E-03 & 56 & Second \\ \hline
215 & IOR.local & 1 & HDF5 Independent N-to-1  File Open Time (Write) & 18.4 & TRUE & 5.60E-02 & 3.42E-03 & 12 & 6.63E-02 & 7.70E-03 & 56 & Second \\ \hline
216 & IOR.local & 1 & HDF5 Independent N-to-1 Read Aggregate Throughput & -1.1 & FALSE & 2.61E+03 & 6.21E+01 & 12 & 2.58E+03 & 3.94E+01 & 56 & MByte per Second \\ \hline
217 & IOR.local & 1 & HDF5 Independent N-to-1 Write Aggregate Throughput & -2.4 & FALSE & 2.90E+01 & 2.05E+00 & 12 & 2.84E+01 & 1.79E+00 & 56 & MByte per Second \\ \hline
264 & IOR.local & 1 & POSIX N-to-N  File Open Time (Read) & -1.9 & FALSE & 4.95E-02 & 5.36E-03 & 12 & 4.85E-02 & 3.67E-03 & 56 & Second \\ \hline
265 & IOR.local & 1 & POSIX N-to-N  File Open Time (Write) & 12.5 & TRUE & 4.86E-02 & 5.50E-03 & 12 & 5.47E-02 & 7.31E-03 & 56 & Second \\ \hline
266 & IOR.local & 1 & POSIX N-to-N Read Aggregate Throughput & -11.3 & TRUE & 2.73E+03 & 3.23E+01 & 12 & 2.42E+03 & 4.63E+02 & 56 & MByte per Second \\ \hline
267 & IOR.local & 1 & POSIX N-to-N Write Aggregate Throughput & -4.1 & TRUE & 9.13E+02 & 5.07E+01 & 12 & 8.76E+02 & 6.76E+01 & 56 & MByte per Second \\ \hline
268 & IOR.local & 1 & Wall Clock Time & 2.2 & FALSE & 4.63E+02 & 1.64E+01 & 12 & 4.73E+02 & 1.90E+01 & 56 & Second \\ \hline
\end{tabular}
\end{table*}

\begin{table*}[!t]
\centering
\footnotesize
\caption{Changes in all measured metrics from MDTest}
\label{table:mdtest-metrics}
\begin{tabular}
{|L{0.1in}|L{0.6in}|R{0.2in}|L{1.0in}|R{0.3in}|R{0.3in}|R{0.4in}|R{0.4in}|R{0.2in}|R{0.4in}|R{0.4in}|R{0.2in}|L{0.6in}|}
\hline
\multirow{2}{*}{\#} & 
\multirow{2}{*}{Application} & 
\multirow{2}{0.2in}{Nodes} & 
\multirow{2}{1.0in}{Metric} & 
\multirow{2}{0.3in}{Diff., \%} & 
\multirow{2}{0.3in}{Distr. Can be Distin- gui- shed?} & 
\multicolumn{3}{l|}{Before Patch Application} & 
\multicolumn{3}{l|}{After Patch Application} &
\multirow{2}{*}{Units} \\[0.1in]  \cline{7-12} 
 &  &  &  &  &  & 
{\centering Mean, Seconds} &l {\centering Standard Deviation, Seconds} & {Nruns} & 
{\centering Mean, Seconds} & {\centering Standard Deviation, Seconds} & {Nruns} & \\[0.1in] \hline
272 & MDTest & 1 & Directory creation (single tree directory) & -15.1 & TRUE & 8.55E+03 & 3.26E+02 & 21 & 7.26E+03 & 3.91E+02 & 56 & Operations/Second \\ \hline
276 & MDTest & 1 & Directory removal (single tree directory) & -16 & TRUE & 2.02E+04 & 7.18E+02 & 21 & 1.70E+04 & 7.99E+02 & 56 & Operations/Second \\ \hline
280 & MDTest & 1 & Directory stat (single tree directory) & -18.4 & FALSE & 1.90E+05 & 7.09E+04 & 21 & 1.55E+05 & 6.58E+04 & 56 & Operations/Second \\ \hline
284 & MDTest & 1 & File creation (single tree directory) & -12.3 & TRUE & 8.09E+03 & 3.13E+02 & 21 & 7.10E+03 & 3.59E+02 & 56 & Operations/Second \\ \hline
288 & MDTest & 1 & File read (single tree directory) & -12.7 & TRUE & 3.28E+04 & 1.96E+03 & 21 & 2.86E+04 & 2.06E+03 & 56 & Operations/Second \\ \hline
292 & MDTest & 1 & File removal (single tree directory) & -15.5 & TRUE & 1.33E+04 & 6.72E+02 & 21 & 1.13E+04 & 5.09E+02 & 56 & Operations/Second \\ \hline
305 & MDTest & 1 & Wall Clock Time & 24.1 & TRUE & 3.05E+01 & 3.17E+00 & 21 & 3.78E+01 & 4.10E+00 & 56 & Second \\ \hline
309 & MDTest & 2 & Directory creation (single tree directory) & -9.8 & TRUE & 4.30E+03 & 2.38E+02 & 23 & 3.88E+03 & 3.44E+02 & 55 & Operations/Second \\ \hline
313 & MDTest & 2 & Directory removal (single tree directory) & -8.5 & TRUE & 3.27E+03 & 2.92E+02 & 23 & 2.99E+03 & 2.68E+02 & 55 & Operations/Second \\ \hline
317 & MDTest & 2 & Directory stat (single tree directory) & -10.3 & TRUE & 2.81E+04 & 1.86E+03 & 23 & 2.53E+04 & 2.25E+03 & 55 & Operations/Second \\ \hline
321 & MDTest & 2 & File creation (single tree directory) & -7.3 & TRUE & 2.25E+03 & 1.08E+02 & 23 & 2.09E+03 & 1.67E+02 & 55 & Operations/Second \\ \hline
325 & MDTest & 2 & File read (single tree directory) & -24.4 & TRUE & 5.94E+04 & 2.90E+03 & 23 & 4.50E+04 & 4.92E+03 & 55 & Operations/Second \\ \hline
329 & MDTest & 2 & File removal (single tree directory) & -10.5 & TRUE & 1.62E+03 & 1.25E+02 & 23 & 1.45E+03 & 9.63E+01 & 55 & Operations/Second \\ \hline
333 & MDTest & 2 & File stat (single tree directory) & -10 & TRUE & 2.19E+04 & 1.62E+03 & 23 & 1.97E+04 & 1.28E+03 & 55 & Operations/Second \\ \hline
342 & MDTest & 2 & Wall Clock Time & 9.7 & TRUE & 1.67E+02 & 3.60E+00 & 23 & 1.83E+02 & 5.30E+00 & 55 & Second \\ \hline
346 & MDTest.local & 1 & Directory creation (single tree directory) & 2 & FALSE & 1.04E+05 & 7.44E+03 & 12 & 1.06E+05 & 8.49E+03 & 56 & Operations/Second \\ \hline
350 & MDTest.local & 1 & Directory removal (single tree directory) & -11.1 & TRUE & 1.30E+05 & 3.70E+03 & 12 & 1.15E+05 & 1.08E+04 & 56 & Operations/Second \\ \hline
354 & MDTest.local & 1 & Directory stat (single tree directory) & -25.6 & TRUE & 1.92E+06 & 5.72E+04 & 12 & 1.43E+06 & 4.68E+04 & 56 & Operations/Second \\ \hline
358 & MDTest.local & 1 & File creation (single tree directory) & -10.9 & TRUE & 1.79E+05 & 1.45E+04 & 12 & 1.59E+05 & 1.54E+04 & 56 & Operations/Second \\ \hline
362 & MDTest.local & 1 & File read (single tree directory) & -25.1 & TRUE & 1.39E+06 & 3.10E+04 & 12 & 1.04E+06 & 2.62E+04 & 56 & Operations/Second \\ \hline
366 & MDTest.local & 1 & File removal (single tree directory) & -8.6 & TRUE & 2.08E+05 & 7.99E+03 & 12 & 1.90E+05 & 1.53E+04 & 56 & Operations/Second \\ \hline
370 & MDTest.local & 1 & File stat (single tree directory) & -26.3 & TRUE & 1.92E+06 & 5.17E+04 & 12 & 1.41E+06 & 2.97E+04 & 56 & Operations/Second \\ \hline
379 & MDTest.local & 1 & Wall Clock Time & 78.6 & TRUE & 3.75E+00 & 6.22E-01 & 12 & 6.70E+00 & 2.61E+00 & 56 & Second \\ \hline
 
\end{tabular}
\end{table*}